\documentclass{aa}

\usepackage{ulem}

\usepackage{color}
\usepackage{array}
\usepackage{url}
\usepackage[bottom]{footmisc}
\usepackage{xcolor}
\usepackage{graphicx}
\usepackage{multirow}
\usepackage{subcaption}
\usepackage{booktabs}
\usepackage{adjustbox}
\usepackage[toc,page]{appendix}
\usepackage{amsmath}
\usepackage{nccmath}
\usepackage{placeins}
\usepackage{capt-of} 
\usepackage{dblfloatfix} 
\usepackage{tikz}
\usetikzlibrary{arrows.meta, bending, shapes.geometric, decorations.pathmorphing, calc}

\usepackage{arydshln} 
\bibpunct{(}{)}{;}{a}{}{,}

\newcommand{\hjdtdb}{\mathrm{HJD}_{\rm TDB}}
\newcommand{\mhjd}{\mathrm{HJD}_{\rm TDB} - 2400000}
\usepackage{txfonts}
 \usepackage{tikz}
 \usetikzlibrary{decorations.pathmorphing, decorations.markings, shapes.geometric, arrows.meta, calc}

\usepackage[]{hyperref}
\hypersetup{unicode=true, colorlinks=true, linkcolor=[rgb]{0.53, 0.15, 0.34}, citecolor=[rgb]{0.0, 0.27, 0.42}, filecolor=[rgb]{1.0, 0.13, 0.32}, urlcolor=[rgb]{0.53, 0.15, 0.34}}

\begin{document} 

  \title{Superorbital variability in the quiescent black hole X-ray transient A0620--00}

\author{Enzo A. Saavedra \inst{1,2} \and
Roi Alonso\inst{1,2} \and
Teo Muñoz-Darias \inst{1,2} \and
Jorge Casares\inst{1,2} \and
Manuel A. P. Torres\inst{1,2} \and \\
Montserrat Armas Padilla\inst{1,2} \and 
James F. Steiner\inst{3} \and
David M. Russell \inst{4} \and 
Gastón J. Escobar\inst{1,2} \and \\
Federico García\inst{5,6} \and 
Fraser Lewis\inst{7,8} \and
Daniel Mata Sánchez\inst{1,2} \and 
Pablo Rodríguez-Gil\inst{1,2}
}

 \institute{
 Instituto de Astrofísica de Canarias (IAC), Vía Láctea s/n, La Laguna 38205, S/C de Tenerife, Spain \and
 Departamento de Astrofísica, Universidad de La Laguna, La Laguna, E-38206, S/C de Tenerife, Spain \and 
 Center for Astrophysics | Harvard \& Smithsonian, 60 Garden St, Cambridge, MA 02138, USA \and
 Center for Astrophysics and Space Science (CASS), New York University Abu Dhabi, UAE  \and
 Instituto Argentino de Radioastronom\'ia (CCT La Plata, CONICET; CICPBA; UNLP), C.C.5, (1894) Villa Elisa, Argentina \and    
 Facultad de Ciencias Astron\'omicas y Geof\'{\i}sicas, Universidad Nacional de La Plata, B1900FWA La Plata, Argentina  \and
 Faulkes Telescope Project, Cardiff, Wales, UK \and
 The Schools' Observatory, Astrophysics Research Institute, Liverpool John Moores University, 146 Brownlow Hill, Liverpool L3 5RF, UK
 }

 \date{Received; accepted}

\abstract{
Quiescent black hole low-mass X-ray binaries provide a key setting for probing accretion physics at low luminosities. A0620--00, the archetypal system in this class, has remained in X-ray quiescence for decades and exhibits complex optical variability, yet the long-term behaviour of its accretion flow remains poorly understood. Here, we report an analysis of long-term optical monitoring of A0620--00 from ZTF, LCO, and ATLAS. The full dataset spans nearly two decades, with the ZTF light curve providing the primary $\sim 2760$-day baseline for the period analysis. We identify a superorbital cycle with a period of $P = 261.9 \pm 9.4$~d and a peak-to-peak amplitude of $\sim 0.2$~mag. The signal is recovered independently across all three surveys, and red-noise simulations indicate that it is unlikely to arise from stochastic variability alone. Furthermore, the relative occurrence of the \textit{passive} and \textit{active} quiescent states displayed by the system seems to depend on the superorbital phase, with passive states concentrated near the cycle minimum and active states more common near maximum. We find that, among the possible interpretations, retrograde nodal precession of a hot inner accretion flow might be able to explain the observed long-term modulation. In this interpretation, the periodic signal may arise from cyclic reorientation of the inner flow, which modulates the photometric contribution from the innermost regions. The inferred modulation period would correspond to a characteristic dynamical radius of $\sim0.13a$ ($\sim10^4~R_{\rm g}$), where $a$ is the binary semi-major axis, broadly consistent with the expected transition between the outer thin disc and the inner hot accretion flow. 
}

 \titlerunning{Superorbital modulation in A0620--00}
 \authorrunning{Saavedra et al.}
 \maketitle

\section{Introduction}
\object{A0620--00} (hereafter A0620) is a well-studied transient black hole low-mass X-ray binary (BH-LMXB) located at a distance of $1.06 \pm 0.12$ kpc \citep{Cantrell2010ApJ...710.1127C}, making it one of the closest known stellar-mass black holes. Since its discovery in outburst in 1975 \citep{Elvis1975Natur.257..656E}, A0620 has remained in X-ray quiescence, enabling detailed dynamical characterisation. The binary comprises a K-type companion star of $0.4~M_\odot$ orbiting a $\sim$6 $M_{\odot}$ BH \citep{McClintock1986ApJ...308..110M, Cantrell2010ApJ...710.1127C, vanGrunsven2017MNRAS.472.1907V}, with an orbital period of 7.75 hours and an orbital decay rate of $\dot{P} \approx -0.60$ ms yr$^{-1}$ \citep{Gonzalez2014MNRAS.438L..21G}.

The optical light curve of A0620 observed during its long-term X-ray quiescence is far from steady, exhibiting complex variability that has been classified into distinct states \citep{Cantrell2008ApJ...673L.159C, Cantrell2010ApJ...710.1127C}. 
The \textit{passive} state is dominated by the ellipsoidal modulation of the donor star, which is produced by the different projected area of the Roche lobe-shaped donor across the orbit \citep[e.g.,][]{vanGrunsven2017MNRAS.472.1907V}. This modulation traces the system’s baseline luminosity. By contrast, A0620 frequently enters an \textit{active} state characterised by aperiodic flaring, an increased mean flux, and significant colour variability.

The sources of optical emission in quiescent BH systems are not fully understood. 
Current evidence suggests that an advection-dominated accretion flow \citep[ADAF;][]{Narayan1994ApJ...428L..13N, Yuan2014ARA&A..52..529Y} governs the inner disc regions at sufficiently low accretion rates \citep[e.g.,][]{McClintock2000ApJ...531..956M, Froning2011ApJ...743...26F}. 
This picture is expected to be particularly relevant when BHs are in quiescence ($L < 10^{-6} L_{\rm Edd}$). 
In the outer regions, the disc is thought to follow the standard geometrically thin, optically thick solution \citep{Shakura1973A&A....24..337S}, and its radiation can be approximated by a multicolour blackbody. 
Thus, the optical emission of A0620 in quiescence includes the companion star together with additional non-stellar components associated with the accretion flow. Donor-subtracted observations further reveal an optical/UV component and a red/infrared excess in quiescence; these may arise in the outer disc or stream--disc impact region, and possibly also from a weak compact jet, circumbinary material, or thermal bremsstrahlung from a warm outflow \citep[e.g.,][]{Froning2007ApJ...663.1215F, Froning2011ApJ...743...26F, Muno2006ApJ...648L.135M, Gallo2007ApJ...670..600G, Cherepashchuk2019MNRAS.483.1067C, Zou2025ApJ...991..157Z}. 
A jet contribution is further supported by the radio and mm detections, as well as by strong near-IR variability and low-level polarization in the reddest bands \citep{Gallo2006MNRAS.370.1351G, Russell2016MNRAS.463.2680R, Dincer2018ApJ...852....4D, Gallo2019MNRAS.488..191G, dePolo2022MNRAS.516.4640D}. 
Variability in these components may contribute to the observed transitions between the different optical quiescent states.

This complex long-term optical behaviour motivates the question of whether part of the variability of A0620 in quiescence may be organised on timescales much longer than the orbital period. 
A possible superorbital modulation of $\sim$255 d has previously been reported in A0620 from long-term optical monitoring, although with only marginal significance \citep{Leibowitz1998MNRAS.300..463L}.
Superorbital flux modulations are well documented in X-rays in high-mass X-ray binaries and in the optical in cataclysmic variables, predominantly in systems accreting at comparatively high rates. Across these systems, the proposed mechanisms for superorbital modulations include disc precession, radiation-driven warping, and variations in the mass-transfer rate \citep{Kotze_Zsources_2010MNRAS.402L..16K, Kotze2012MNRAS.420.1575K, Fragile2026arXiv260302993F}.
In LMXBs, convincing cases are rarer, and the clearest examples are found in neutron star systems. These include ultracompact X-ray binaries (i.e. with $P_{\rm orb} \lesssim 80$~min; e.g., \citealt{ArmasPadilla2023A&A...677A.186A}), such as \object{4U~1820--303} and \object{47~Tuc~X--9} \citep{Richman1994PASP..106.1075R, Tudor2018MNRAS.476.1889T}, as well as other neutron star LMXBs such as \object{4U~1636--53} \citep{Shih2005MNRAS.361..602S}.
However, such cases remain rare, and in BH-LMXBs these modulations have generally been reported in bright sources \citep[see e.g.,][]{Kotze2012MNRAS.420.1575K}.  
This difficulty is not only a matter of the typically longer orbital periods of BH-LMXBs, but also of the fact that they spend most of their lifetimes in quiescence.

In an attempt to address this gap, we use optical monitoring from different facilities to search for long-term modulations in A0620. We describe the observational data set in the following section; in Section~\ref{sec:results} we analyse and evaluate the significance of a 262~d periodicity identified in these data, and in Section~\ref{sec:discuss} we discuss possible origins of this signal.

\section{Photometric data} \label{sec:photdata}

Long-term optical photometry of A0620 was retrieved from the Zwicky Transient Facility (ZTF), the Las Cumbres Observatory (LCO) and the Asteroid Terrestrial-impact Last Alert System (ATLAS). A summary of these observations, including the temporal coverage and number of data points for each facility, is presented in Table~\ref{tab:obs_log}. The ZTF dataset, spanning $\sim$2800~d, provides the primary baseline for our period analysis.

\subsection{Zwicky Transient Facility}

A0620 was observed using the ZTF, a time-domain survey based on the Palomar 48-inch Schmidt Telescope \citep{Masci2019PASP..131a8003M}. ZTF features a 600-megapixel camera with a 47 deg$^2$ field of view, surveying the northern visible sky at $\sim3760$ deg$^2$ h$^{-1}$ and achieving median depths of $g \approx 20.8$ and $r \approx 20.6$ (AB; $5\sigma$ in 30 s).  The A0620 dataset released in DR24 spans $\mhjd = 58204$--60968, comprising 348 and 334 epochs in the $r$- and $g$-bands, respectively. The ZTF pipeline provided photometric and astrometric calibration with relative photometric accuracy $\lesssim2\%$ versus Pan-STARRS DR1. Simultaneous data from four comparison stars within the same field were included (see Fig.~\ref{fig:fov} and Table~\ref{tabla:mag_2dp}). The ZTF monitoring covers a total baseline of $\sim 2764$ d, structured into seven seasonal observing windows with an average duration of $\sim 206$ d.

\begin{table}[h!]
\caption{Log of Photometric Observations.}
\label{tab:obs_log}
\centering
\resizebox{\columnwidth}{!}{%
\begin{tabular}{l c c c r}
\toprule
Telescope & Filter & $\hjdtdb$ - 2400000 Range & Span (d) & $N_{\rm obs}$ \\
\midrule
ZTF & $g$ & 58204--60968 & 2764 & 334 \\
ZTF & $r$ & 58216--60968 & 2752 & 348 \\
LCO & $V$ & 53750--60740 & 6988 & 523 \\
LCO & $i$ & 53750--60740 & 6989 & 651 \\
ATLAS & $o$ & 57318--60540 & 3222 & 2420 \\
ATLAS & $c$ & 57313--60400 & 3087 & 690 \\
\bottomrule
\end{tabular}%
}
\tablefoot{
The $\hjdtdb$ - 2400000 range is rounded to the nearest integer.
}
\end{table}

\subsection{Las Cumbres Observatory}

A0620 has been regularly monitored in the optical during the last $\sim20\ \mathrm{yr}$ with the LCO 2\,m and 1\,m robotic telescopes, from 2006 January 15 (HJD$_{\rm TDB}$ = 2453750) to 2025 March 3 (HJD$_{\rm TDB}$ = 2460740), mostly using $V$ and $i$ filters. The images have been processed and analysed by the recently developed \textsc{xbnews} pipeline, which downloads the reduced images (i.e., bias, dark, and flat-field corrected images) from the LCO archive, automatically rejects poor-quality reduced images, performs astrometry using \textit{Gaia} DR2 positions and carries out multi-aperture photometry \citep[MAP;][]{Stetson1990PASP..102..932S}, solves for photometric zero-point offsets between epochs \citep{Bramich2012MNRAS.424.1584B}, and flux-calibrates the photometry using the ATLAS-REFCAT2 catalogue \citep{Tonry2018ApJ...867..105T}. 
If the target is not detected in an image above the detection threshold, then \textsc{xbnews} performs forced MAP at the target coordinates. In this case, we reject all forced MAP magnitudes with an uncertainty $>0.25\ \mathrm{mag}$, as these are very uncertain photometric measurements. For further details on \textsc{xbnews}, see \citet{Russell2019AN....340..278R} and \citet{Goodwin2020MNRAS.498.3429G}. For further details of the LCO/Faulkes monitoring of A0620, see \citet{Zou2025ApJ...991..157Z}. Up to this analysis cutoff date, a total of 651 and 523 reliable magnitudes in the $i$ and $V$ bands, respectively, were obtained during our ongoing long-term optical monitoring of A0620 with LCO.

\subsection{Asteroid Terrestrial-impact Last Alert System}

A0620 was also observed with ATLAS \citep{Tonry2018ApJ...867..105T}, which consists of four 0.5 m telescopes located in Hawaii, Chile, and South Africa. ATLAS surveys a large fraction of the sky with approximately a daily cadence. ATLAS' telescopes employ customised filters \textit{cyan} ($c$; covering 420--650 nm) and \textit{orange} ($o$; covering 560--820 nm). Forced photometry\footnote{https://fallingstar-data.com/forcedphot/} of A0620 was conducted from $\mhjd = 57313$--60540 using photometry measured on the ATLAS reduced images. Flux calibration relied on the ATLAS Refcat2 catalogue. 

\begin{figure}[!h]
\centering
\includegraphics[width=\linewidth]{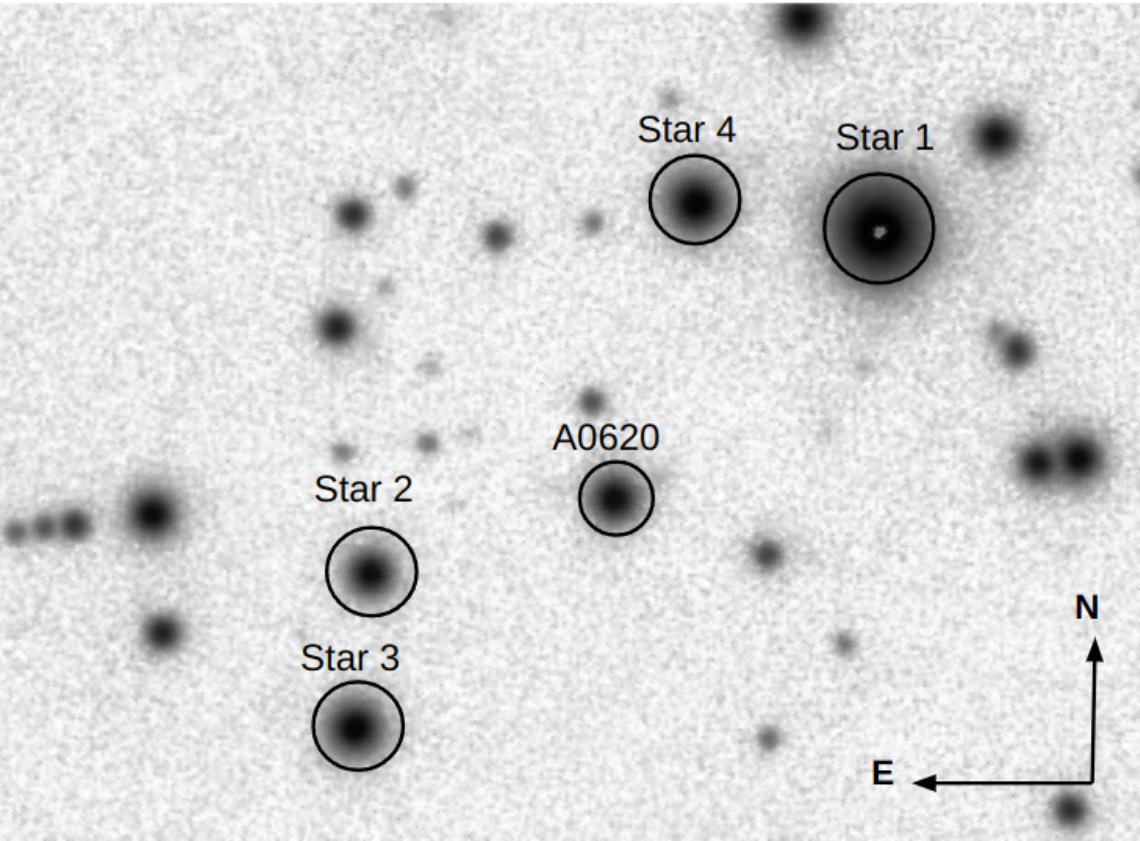}
\caption{Pan-STARRS DR1 (i, r, g) image showing A0620 and the comparison stars used in the ZTF data. The field of view is $\sim1'\times1'$.}
\label{fig:fov}
\end{figure}

\begin{figure*}
  \centering
  \includegraphics[width=0.45\linewidth]{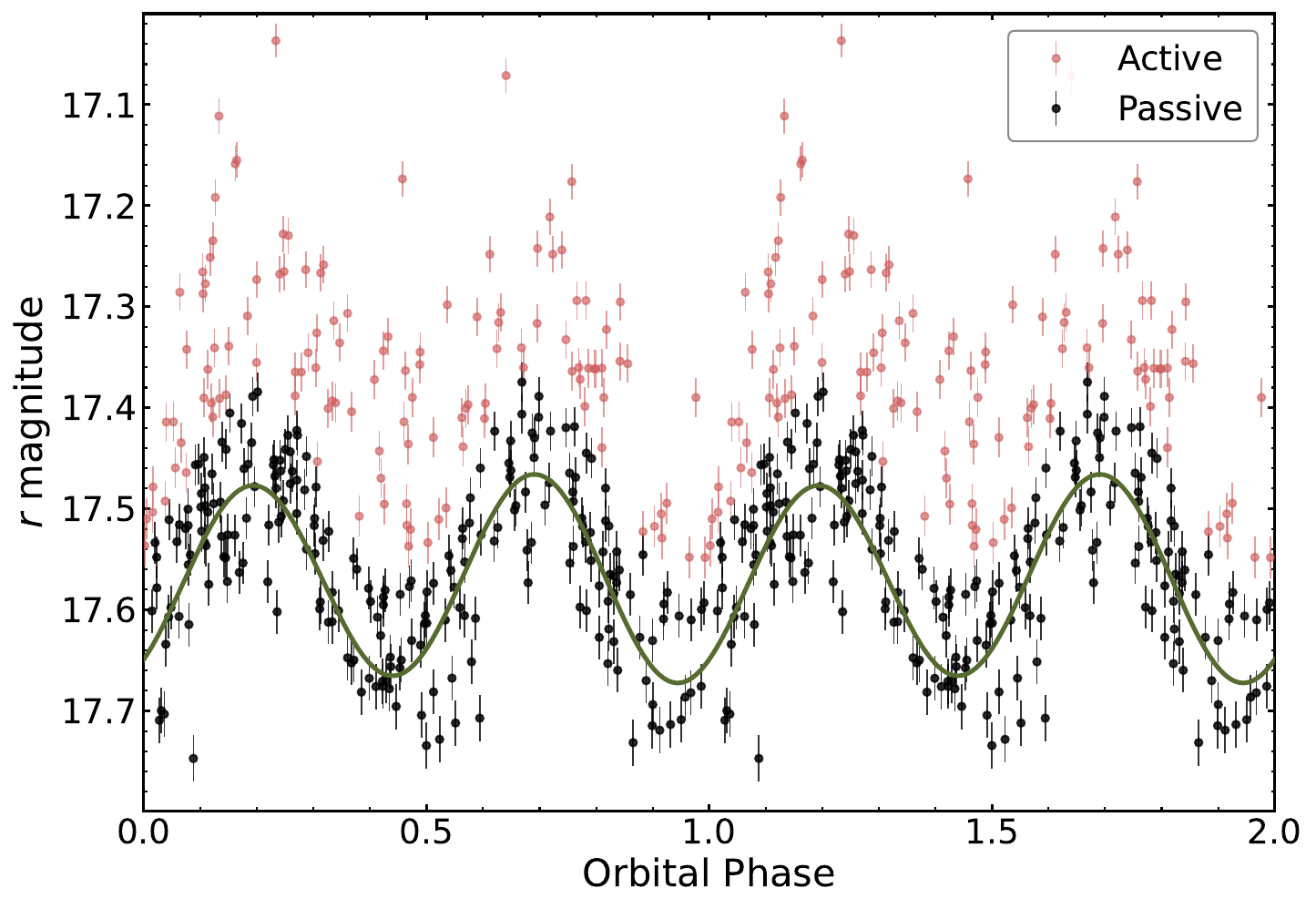} \includegraphics[width=0.45\linewidth]{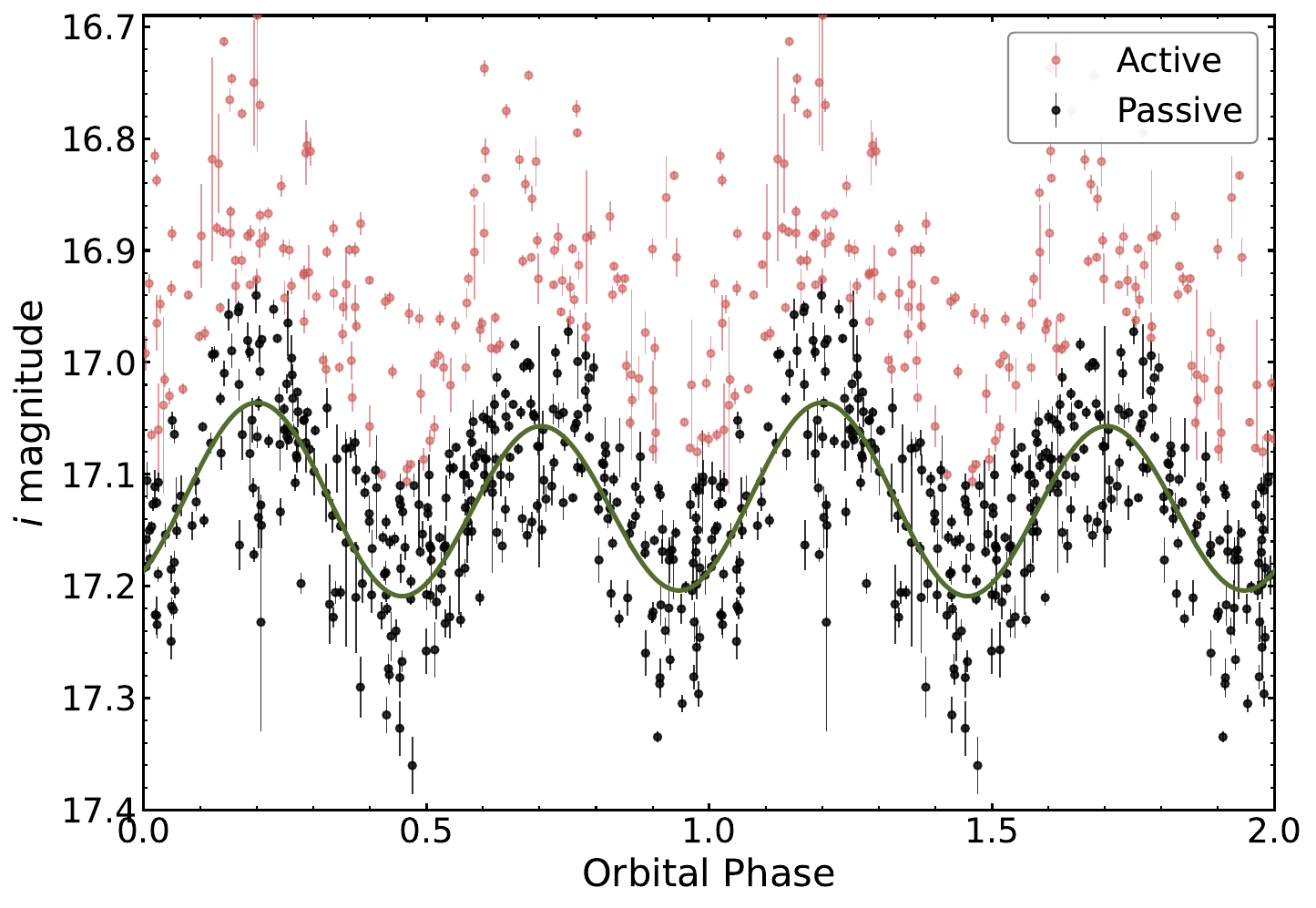} 
 \caption{ZTF-$r$ (left) and LCO-$i$ (right) light curves of A0620, folded at the orbital period. The plots illustrate the ellipsoidal modulation, distinguishing between active (red) and passive (black) states. The green curve shows the Fourier model used to define the passive baseline.}
  \label{fig:ellipsoidal}
\end{figure*}

\section{Analysis and Results} \label{sec:results}

We first characterised the long-term optical variability of A0620 across the available datasets.
For each survey and filter we computed the mean magnitudes and root-mean-square (rms) dispersions, obtaining $r = 17.5 \pm 0.1$ mag and $g = 18.6 \pm 0.1$ mag from ZTF, $o = 17.4 \pm 0.2$ mag and $c = 18.1 \pm 0.2$ mag from ATLAS, and $V = 17.6 \pm 0.3$ mag and $i = 17.1 \pm 0.1$ mag from LCO.

To assess whether the variability is intrinsic to A0620 rather than an instrumental artefact, we selected comparison stars from the ZTF dataset with high-quality photometry (Fig.~\ref{fig:fov}; Table~\ref{tabla:mag_2dp}). The comparison stars show low rms scatter, with $\sigma_{\rm g}=0.024$--$0.066$~mag and $\sigma_{\rm r}=0.026$--$0.053$~mag (Table~\ref{tabla:mag_2dp}). The rms dispersion in the A0620 photometry is $\sim2.5$ times larger than the scatter observed in a comparison star with similar brightness, supporting the conclusion that the excess scatter is intrinsic to A0620.

As a first step, we performed a photometric classification following \citet{Cantrell2010ApJ...710.1127C} into passive and active states (see Fig.~\ref{fig:ellipsoidal}). We implemented this separation by modelling the lower envelope of the phase-folded light curve, using the orbital period and reference epoch \citep[7.75234 h and HJD$_{\rm TDB}$ = 2446082.6671;][]{Gonzalez2014MNRAS.438L..21G}. Specifically, we fitted a second-order Fourier series in orbital phase to the faint end of the magnitude distribution (defined by the 75th percentile in phase bins) to establish the stable passive baseline; this model is overplotted in the phased light curve for reference. We then computed the residuals of the full dataset relative to this model.
Data consistent with the baseline (residuals $\lesssim 0.1$ mag, allowing for photometric noise) were classified as passive, while those exhibiting significant excess brightness were classified as active (see Fig.~\ref{fig:ellipsoidal}). During these active intervals, the system brightens by $\Delta m \approx 0.3$ mag relative to the passive state across different orbital phases (e.g., from $r \approx 17.4$ to $17.1$ at phase 0.25, and from $17.5$ to $17.2$ at phase 0.5), implying that the excess emission accounts for at least $\sim 25\%$--$30\%$ of the total optical flux.
This classification approach yields a physically motivated distinction between intervals governed by the ellipsoidal modulation of the donor star (i.e., passive state) and those dominated by aperiodic variability (i.e., active state).

\begin{figure*}
  \centering
  \includegraphics[width=0.4925\linewidth]{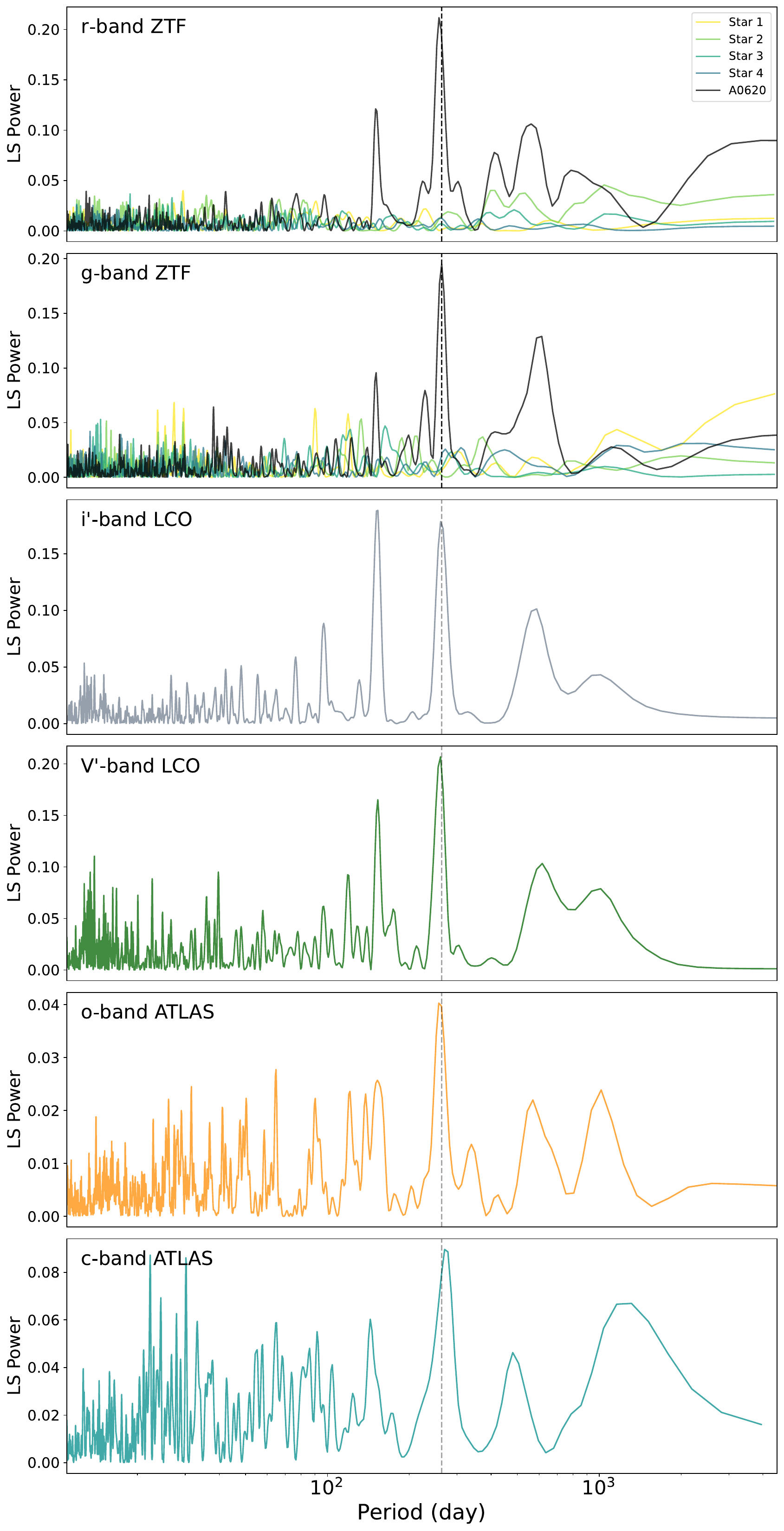}
  \includegraphics[width=0.495\linewidth]{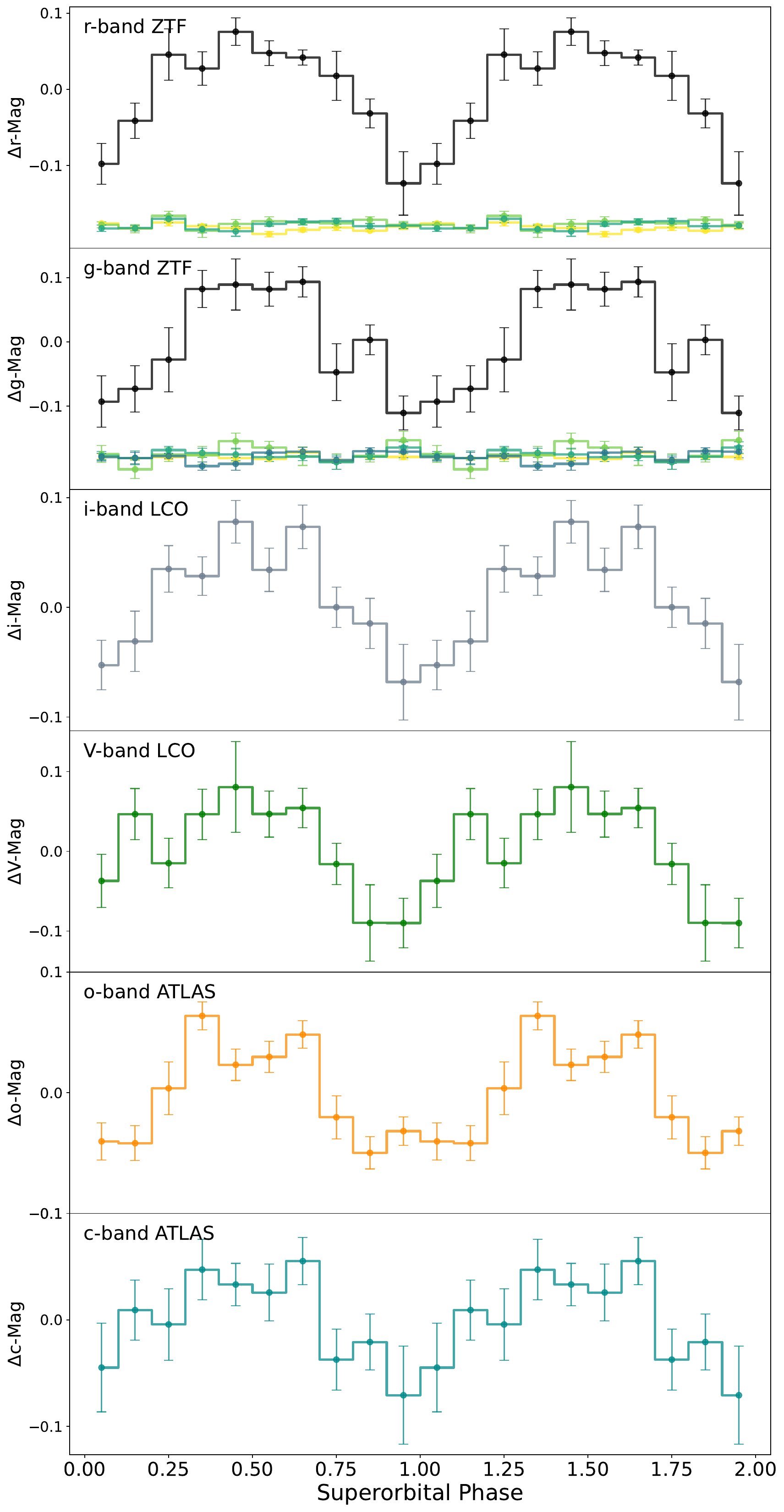}
  \caption{Left: LS periodograms of A0620 from multi-survey photometry. The upper panels show ZTF $g$- and $r$-band power spectra (comparison stars included). The lower panels present LCO ($i$-, $V$-band) and ATLAS ($o$- and $c$-band) periodograms. The 262-d periodicity (vertical dashed line) appears consistently across all datasets. Right: Corresponding phase-folded light curves at $P = 262$ d. ZTF $g$- and $r$-band data (top panels) include comparison stars offset by 0.18 mag. ATLAS and LCO data (bottom four panels) show coherent modulation across all bands, supporting the periodicity.}
  \label{fig:lombscargles}
\end{figure*}

\begin{table}[!h]
\centering
\caption{Mean ZTF magnitudes of A0620 and comparison stars.}
\label{tabla:mag_2dp}
\begin{tabular}{lcccc}
\toprule
Star  & $g$ mag      & $\sigma_{\rm g}$ (mag) & $r$ mag      & $\sigma_{\rm r}$ (mag) \\
\midrule
1   & $15.25$ & 0.024      & $14.63$ & 0.026 \\
2   & $18.56$ & 0.066      & $17.63$ & 0.053 \\
3   & $17.87$ & 0.049      & $17.07$ & 0.033 \\
4   & $17.63$ & 0.048      & $16.69$ & 0.037 \\
A0620 & $18.59$ & 0.166      & $17.47$ & 0.134 \\
\bottomrule
\end{tabular}
\tablefoot{Mean $g$- and $r$-band magnitudes and light-curve rms dispersions are given for A0620 and the four comparison stars in the ZTF dataset. The $\sigma$ columns list the light-curve rms dispersion.}
\end{table}

\subsection{Long-term periodicity}

In light of the observed variability, we computed single-band Lomb--Scargle (LS) periodograms independently for each filter on the full A0620 light curve, that is, including both the passive and active states, see Fig.~\ref{fig:lombscargles} \citep{Lomb1976Ap&SS..39..447L, Scargle1982ApJ...263..835S}. 
We restricted the LCO and ATLAS observations to epochs after HJD$_{\rm TDB}$ = 2458212, to maximise cross-survey overlap and avoid calibration changes at earlier epochs (see below). 
Each dataset showed that the highest-power peak corresponds to a period of $\sim$262\,d, with additional peaks at $\sim$155\,d (likely the first harmonic) and $\sim$643\,d (likely a sub-harmonic of the main signal). These periodicities were confirmed by additional analysis using phase dispersion 
minimisation \citep{Stellingwerf1978ApJ...224..953S} and the non-uniform fast Fourier transform method \citep{lombscargel2024arXiv240908090G}. 
The distributions of the LS period estimates are adequately described by a Gaussian; we therefore combine the band-by-band period estimates through an inverse-variance weighted mean, using their $1\sigma$ statistical uncertainties to define the weights (Table~\ref{periods}). Propagating these statistical errors into the combined, band-averaged period yields $P = 261.9 \pm 8_{\mathrm{stat}}$ d.

\begin{figure}
  \centering
  \includegraphics[width=1\linewidth]{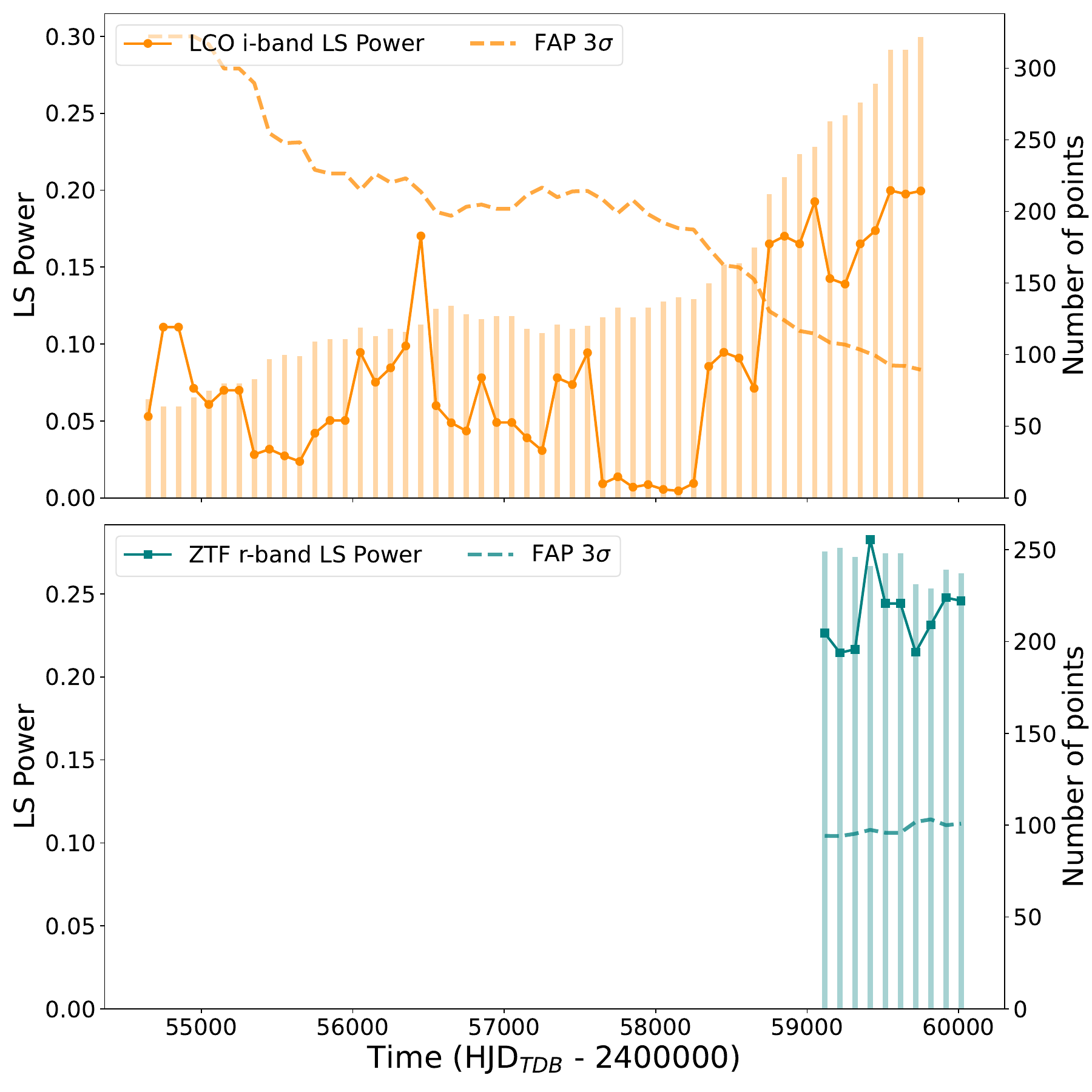}
  \caption{Periodogram analysis as a function of time. LS power at 262 d (filled circles), the $3\sigma$ false-alarm threshold (dashed lines) and the number of observations per segment (shaded bars). }
  \label{fig:power_LCO}
\end{figure}

\begin{figure*}[!t]
  \centering
  \includegraphics[width=0.95\linewidth]{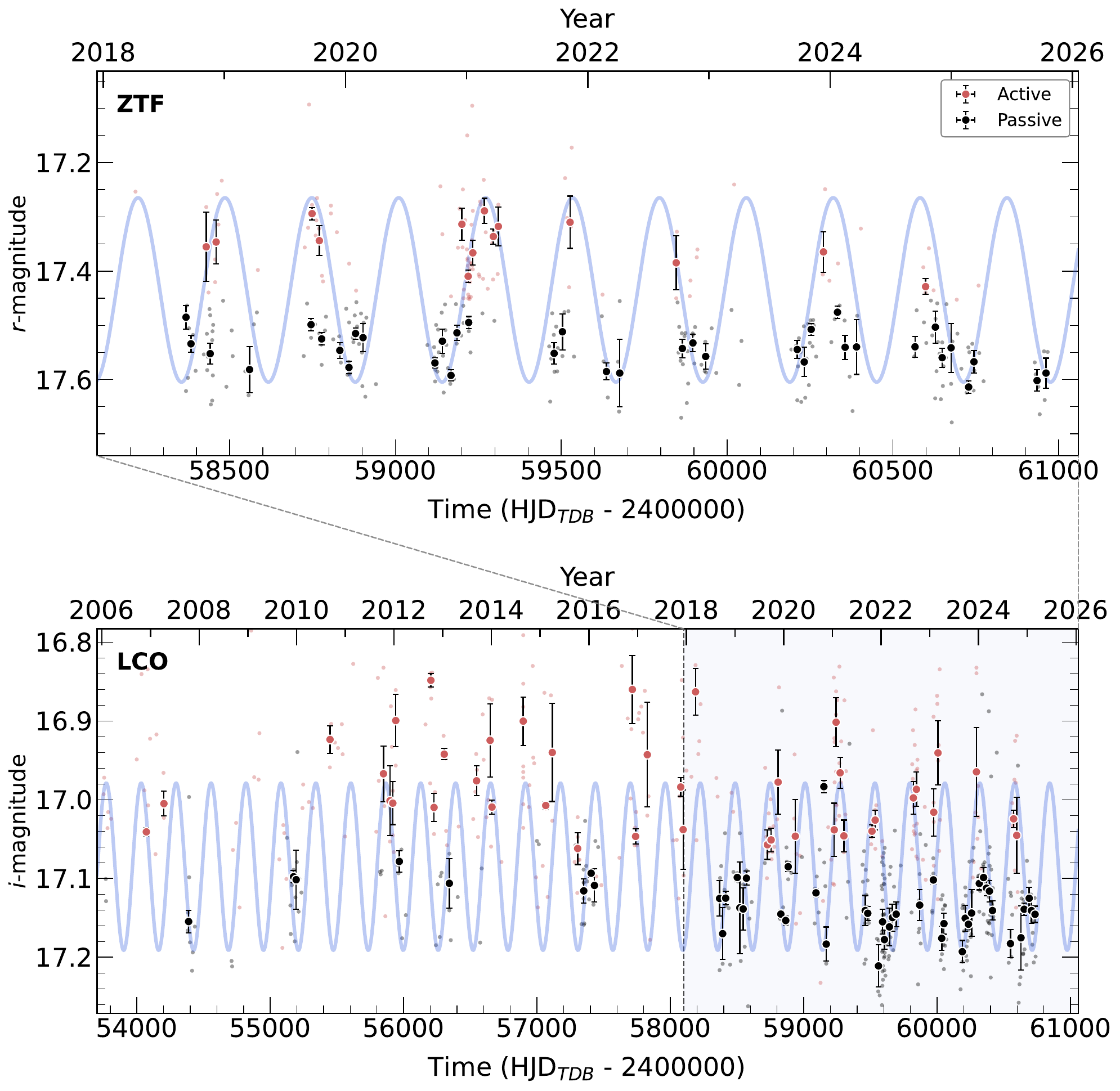}
  \caption{Light curves of A0620 covering the long-term modulation. \textit{Top:} ZTF $r$-band. \textit{Bottom:} LCO $i$-band. \\
  The panels cover different timescales, with lines indicating the zoom-in from the ZTF baseline to the LCO period. In both panels, the faint background points show the individual observations, while the larger points show the median in 26-day bins, with error bars derived from the median absolute deviation. A sinusoidal model with a period of 262 days is overplotted in each panel, and all points are colour-coded according to the source state.}
  \label{fig:binned-rband}
\end{figure*}

To account for band-to-band differences and for reasonable variations in time span and state selection, we further include a systematic contribution to the period uncertainty, estimated from the scatter between the periods obtained in the different filters and from these alternative analyses, which contributes $\sigma_{P,\mathrm{sys}} = 5$ d and is added in quadrature to the statistical term\footnote{We define the systematic uncertainty as the sample standard deviation of the set of periods $\{P_i\}$ obtained from the different bands and alternative analyses, $\sigma_{P,\mathrm{sys}} = \left[\sum_{i=1}^{N} (P_i - \bar{P})^2/(N-1)\right]^{1/2}$, where $\bar{P}$ is their mean and $N$ is the number of measurements. The total uncertainty is then $\sigma_{P,\mathrm{tot}} = (\sigma_{P,\mathrm{stat}}^2 + \sigma_{P,\mathrm{sys}}^2)^{1/2}$.}.
Our final constraint on the long-term modulation is therefore $P = 261.9 \pm 8_{\mathrm{stat}} \pm 5_{\mathrm{sys}} = 261.9 \pm 9.4~\mathrm{d}$
(234--290~d with $3\sigma$ confidence). 
The peak-to-peak variations are consistent with $\sim 0.2$\,mag across all cases. Semi-amplitudes show no significant dependence on passband; for instance, the $g$-band semi-amplitude is $0.10 \pm 0.01$, comparable to $0.085 \pm 0.009$ in the $r$ band.

To define a common superorbital ephemeris, we fitted each band independently with a sinusoid of fixed period $P=261.9$ d, leaving the mean level, amplitude, and phase free. We define $T_0$ as the epoch of maximum magnitude of the modulation, corresponding to the faint phase of the cycle. The fitted $T_0$ values cluster around HJD$_{\rm TDB}$ = 2459520, although modest band-to-band offsets remain. Because the single-sinusoid fits show excess scatter relative to the formal photometric uncertainties, we rescaled the phase uncertainties to account for the intrinsic variability before combining the measurements. The resulting representative common epoch is $T_0[\mathrm{HJD}_{\rm TDB}] = 2459520 \pm 6$. This value is adopted as the common reference epoch throughout this work.

\subsection{On the Significance of the Periodicity}

To quantify the confidence in this periodicity, we assessed its statistical significance by calculating the false alarm probability (FAP) for each peak and converting it to $\sigma$ significance with the inverse standard normal survival function using the {\tt astropy} library \citep{Astropy2022ApJ...935..167A}. The $\sim262$-day peak reaches equivalent single-trial Gaussian significances of $6.6\sigma$ in the $r$ band and $5.8\sigma$ in the $g$ band. The ATLAS $o$ and $c$ bands give significances of $10\sigma$ and $4.6\sigma$, respectively. While the LCO data also indicate the presence of the 262~d periodicity, it remains below the $3\sigma$ threshold; this is plausibly due to the smaller effective baseline/cadence and larger photometric error in the LCO sample relative to ZTF/ATLAS. 

We tested whether the 262-day peak could arise from red noise. We fitted the log–log power-spectrum continuum (masking the 262~d peak and its harmonics), obtaining power-law slopes $\alpha_{\rm r}=0.35\pm0.07$ and $\alpha_{\rm g}=0.32\pm0.15$ for the $r$ and $g$ bands, respectively. 
Using these slopes, we simulated $2\times10^{5}$ red-noise light curves with the actual sampling and photometric errors \citep[see e.g.,][]{Timmer1995A&A...300..707T, Saavedra2025A&A...702A.103S}, computed LS periodograms, and recorded the maximum power over 10--1000\,d. In $r$, the power at $P=262\,\mathrm{d}$ lies in the extreme tail of the simulated distribution, corresponding to a global significance of $\simeq5\sigma$ (and $\simeq4.8\sigma$ in $g$). Given the finite number of simulations, we refrain from quoting a more precise value, but the result shows that the $\sim262$~d modulation is highly unlikely to be produced by red noise alone.

To assess potential contamination from residual differential photometry effects, we computed the periodogram of airmass as a function of observing time. This periodogram shows pronounced peaks at annual and semi-annual periods, consistent with seasonal visibility variations, whereas the periodogram of A0620 does not exhibit these features. The absence of these peaks suggests that chromatic effects do not dominate the observed periodicity, reinforcing the conclusion that the $\sim262$ d signal is intrinsic to A0620.
To verify that ellipsoidal modulation did not contribute to the 262-day periodicity, we sampled the ellipsoidal model at the same observation epochs and computed its LS periodogram. No significant periodicity emerged. This indicates that the 262-day signal does not result from the continuous but partial sampling of the underlying ellipsoidal modulation. Having established this, we subtracted the best-fit ellipsoidal model from the observed light curves. The 262-day signal persists in the residuals with consistent amplitude, and hereafter all analyses are performed on these ellipsoidal-subtracted light curves unless stated otherwise.

To investigate intermittent signals and the evolution of the $\sim$262-day modulation, we computed moving Lomb-Scargle (LS) periodograms \citep{Kalicinsky:2020aa, VanderPlas2018ApJS..236...16V}. We monitored general source behaviour by mapping signal power and amplitude within a 1000-day sliding window. The results of this detectability analysis show that in early LCO epochs (HJD$_{\rm TDB} \lesssim 2458750$), where sampling is sparse ($N \simeq 70$--175), the LS power remains below the threshold (see Fig.~\ref{fig:power_LCO}). We attribute this non-detection to the combination of poor phase coverage and stochastic noise contamination, which effectively raises the noise floor. However, as cadence improves after HJD$_{\rm TDB} \sim 2458750$ ($N \gtrsim 210$), the modulation becomes significantly detected. The ZTF $r$-band data, with consistently denser sampling ($N \simeq 250$), exceeds the threshold in all windows, supporting that the signal is persistent but requires sufficient cadence to emerge from the stochastic noise in the active states.

To explore whether these early non-detections might be primarily a sampling artifact, we performed a Monte Carlo rejection test. We selected the most densely sampled 1800-day window in the LCO dataset ($N = 322$) and randomly downsampled it to match the observation count of our sparsest early window ($N = 64$). Across $10^4$ iterations, we found that in $\approx$15.4\% of the trials, this artificial degradation produced an LS power equal to or weaker than the actual value observed during the early epochs. This result indicates that the apparent absence of the modulation before HJD$_{\rm TDB} \sim 2458750$ is statistically compatible with sparse sampling.

\begin{table}
  \centering
  \caption{Periods obtained from each band periodogram.}
  \begingroup
  \small 
  \setlength{\tabcolsep}{7pt}
  \begin{tabular}{@{}l*{6}{c}@{}}
    \toprule
     & $r$ & $g$ & $i$ & $V$ & $o$ & $c$ \\
    \midrule
    Period [d] & 264.8 & 262.1 & 263.4 & 258.9 & 260.4 & 268.9 \\
    Error [d] & 18.3 & 19.8 & 12.7 & 11.4 & 15.4 & 24.9 \\
    \bottomrule
  \end{tabular}
  \label{periods}
  \endgroup
  \tablefoot{Periods and errors $(1\sigma)$ are given in units of days. The periodograms are associated with each band for the epochs ranging from MJD 58212 to 59998.}
\end{table}

\subsection{Phase-Folded Light Curves}

We folded the light curves using the superorbital ephemeris defined above, adopting $P=261.9$ d and $T_0[\mathrm{HJD}_{\rm TDB}] = 2459520$, and calculated median magnitudes in 0.1-phase bins. Fig.~\ref{fig:lombscargles} (right panels) shows a similar modulation in A0620 across all bands when the data are phase-folded on this ephemeris. In contrast, the comparison stars remain statistically constant under identical analysis, supporting that the periodic variability is intrinsic to A0620.

In an effort to better visualise the long-term periodicity, the ZTF $r$-band and LCO $i$-band were binned into 26 d intervals. The median was computed for each bin and a sinusoidal model with a period of 262 days was plotted (see Fig.~\ref{fig:binned-rband}). Although significant residuals are evident, the 262-day modulation broadly traces the long-term evolution of the system flux,
with the maxima roughly corresponding to epochs dominated by the active state and the minima to passive periods.

\begin{figure}
  \centering
  \includegraphics[width=0.9\linewidth]{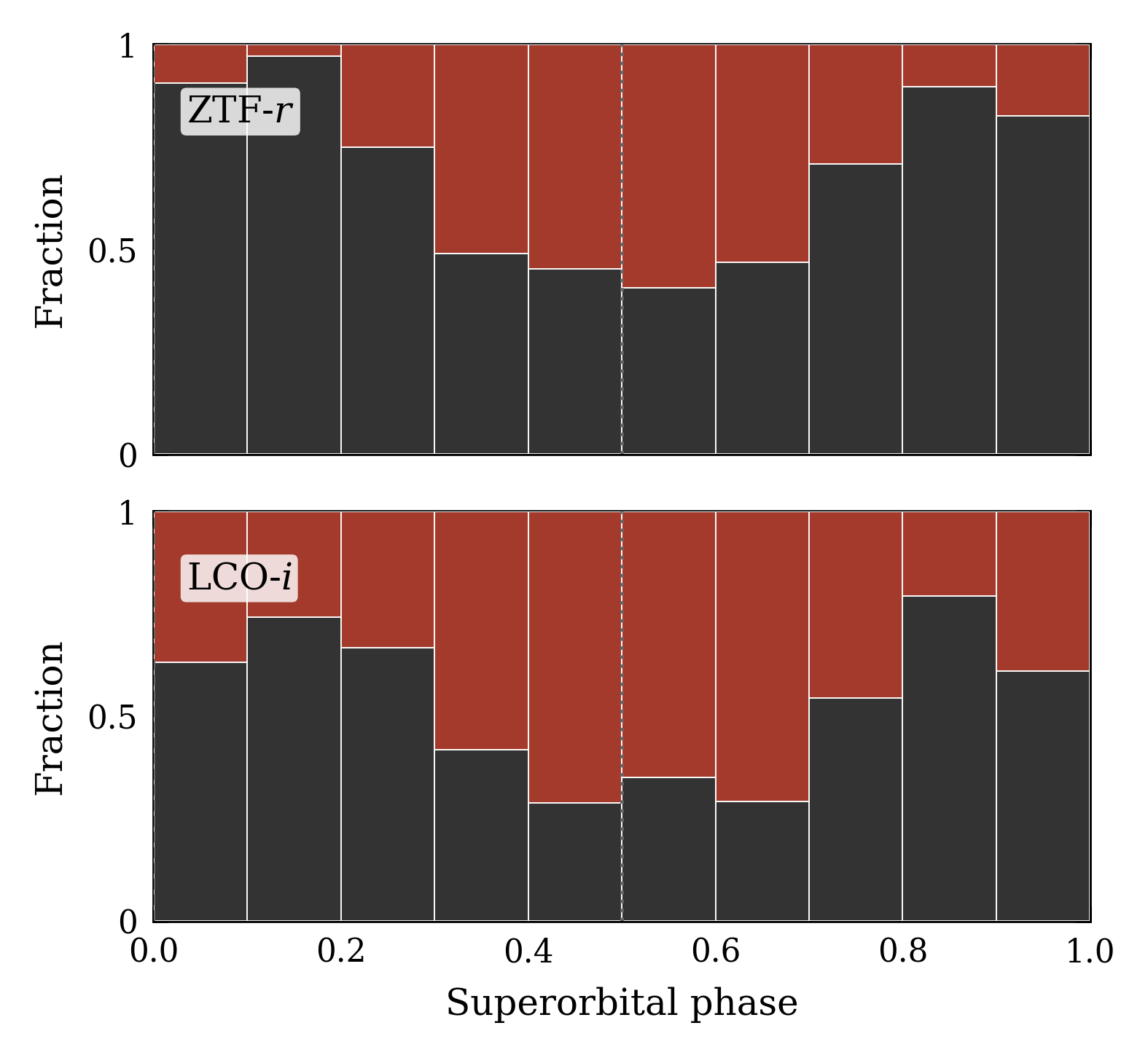}
\caption{
  Phase histograms of the passive and active classifications. Black denotes the passive fraction and red denotes the active fraction. The histograms are normalised by the total number of observations in each of ten equal superorbital phase bins.  The upper and lower panels correspond to the ZTF-$r$ and LCO-$i$ data, respectively.
}
\label{fig:state_hist}
\end{figure}

To further understand the connection between the 262-day variability and the quiescent-state classification, we computed histograms of the passive and active data points in ten equal superorbital phase ($\phi_{\rm sup}$) bins and obtained their phase-dependent fractions (see Fig.~\ref{fig:state_hist}).
In the ZTF-$r$ data, the $\phi_{\rm sup}=0.0$--$0.1$ bin contains 29 passive and 3 active points, corresponding to passive and active fractions of 91\% and 9\%, respectively. By contrast, the $\phi_{\rm sup}=0.5$--$0.6$ bin contains 13 passive and 19 active points, corresponding to 41\% and 59\%, respectively. Around $\phi_{\rm sup}\simeq0.5$, the passive fraction therefore drops substantially, while the active fraction increases.
A similar behaviour is seen in the LCO-$i$ data. The $\phi_{\rm sup}=0.0$--$0.1$ bin contains 35 passive and 19 active points, corresponding to passive and active fractions of 65\% and 35\%, respectively. In the $\phi_{\rm sup}=0.5$--$0.6$ bin, these numbers change to 21 and 40, corresponding to 34\% and 66\%. 
Around $\phi_{\rm sup}\simeq0.5$, the passive fraction therefore drops while the active fraction increases, a behaviour seen across both datasets.

\section{Discussion} \label{sec:discuss}

In this work, we report evidence for superorbital variability in A0620. We detect a superorbital modulation at $261.9\pm9.4$\,d ($>5\sigma$) with a peak-to-peak amplitude of $\sim0.2$\,mag. The feature is recovered independently in the ZTF, ATLAS, and LCO datasets, making instrumental or survey-specific systematics improbable. Since the comparison stars show no analogous behaviour, these results suggest that the periodicity is likely intrinsic to A0620.

A more detailed inspection shows that the $\sim262$-day modulation is most clearly recovered in the ZTF dataset over the interval $\mhjd \simeq 58700$--$60000$, corresponding to roughly five cycles. The same feature is also seen in the ATLAS data over a similar time span. In the longer-baseline LCO data, however, the periodicity is recovered less uniformly; as shown by the moving periodograms (see Fig.~\ref{fig:power_LCO}), it is more prominent during some epochs and less evident during others. In particular, the signal is weaker prior to HJD$_{\rm TDB} \sim 2458700$.

It remains unclear whether this reduced visibility arises from changes in the intrinsic behaviour of the source, sparse sampling in the earlier part of the LCO light curve, additional stochastic noise, or some combination of these effects. Although we cannot rule out genuine long-term intermittency, the behaviour seen during the better-sampled intervals, together with the overall consistency across datasets, suggests that sparse sampling may be the dominant explanation for the apparent weakening of the feature in the early LCO data. Even so, the overall behaviour remains consistent with the $\sim$255\,d cycle reported by \citet{Leibowitz1998MNRAS.300..463L}, despite the marginal significance of that earlier detection. The fact that two independent analyses, separated by nearly three decades, recover similar periods lends further support to the presence of a long-term cycle in A0620.

The phase histograms show that the relative fractions of the two quiescent-state classifications vary with superorbital phase, with passive points more frequent near $\phi_{\rm sup}\simeq0$ and active points near $\phi_{\rm sup}\simeq0.5$.
This indicates that the $\sim$262-day modulation is linked not only to the long-term flux evolution, but also to the changing relative occurrence of the two quiescent states. By definition, during intervals classified as active, individual data points frequently lie well outside the sinusoidal trend, with deviations of up to $\sim$0.2 mag (Fig.~\ref{fig:ellipsoidal}). This is comparable to the peak-to-peak amplitude of the long-term cycle itself (see Fig.~\ref{fig:binned-rband}). This behaviour is consistent with the superposed aperiodic variability that characterises the active state \citep{Cantrell2008ApJ...673L.159C, Cantrell2010ApJ...710.1127C}, which has been attributed to the inner accretion flow and/or the jet \citep{Shahbaz2004MNRAS.354...31S, Veledina_2013, Gallo2006MNRAS.370.1351G, Gallo2019MNRAS.488..191G, Russell2016MNRAS.463.2680R}. We now consider the physical mechanisms that could account for both the long-term modulation and its connection to the state-dependent optical variability in A0620.

\subsection{Nodal Precession of the Hot Inner Flow}

A plausible scenario is the retrograde \textit{nodal precession} of the accretion flow, driven by the tidal torque of the companion star on a misaligned disc structure. In this scenario, the angular-momentum vector of the tilted flow precesses retrogradely around the binary angular-momentum vector due to the companion’s gravitational pull. This is a dynamical effect analogous to the nodal precession of the Moon's orbit, which is tilted with respect to the Earth's orbital plane and precesses under the influence of the Sun's gravity. This gradual change in disc orientation produces a periodic modulation in the observed emission that can naturally reach timescales of hundreds of days. Disc precession of this kind has been proposed to explain the long-term evolution observed in several X-ray binaries, including the persistent neutron-star system \object{Her~X-1}, with its well-known $\approx 35$\,d cycle \citep{Katz1973NPhS..246...87K, Gerend1976ApJ...209..562G, Wijers1999MNRAS.308..207W}.

To assess whether this interpretation is dynamically plausible, we apply the formalism for a fluid disc subject to binary torques \citep{Papaloizou1995MNRAS.274..987P,Larwood1996MNRAS.282..597L,Bate2000MNRAS.317..773B}. Assuming the precessing structure behaves as a rigid body, the relationship between the precession period $P_{\rm prec}$ and the orbital period $P_{\rm orb}$ is described by:

\begin{equation}
  \left| \frac{P_{\rm orb}}{P_{\rm prec}} \right| = \frac{15}{32} \frac{q}{\sqrt{1+q}} \left( \frac{R_{\rm d}}{a} \right)^{3/2} \cos \delta,
\end{equation}

\noindent where $q = M_2/M_1$ is the mass ratio, $a$ is the binary separation, and $\delta$ is the tilt angle (assumed small, $\cos \delta \approx 1$). Substituting the system parameters for A0620 ($P_{\rm orb} \approx0.323$~d, $P_{\rm prec} \approx 261.9$~d, $q\approx0.06$) into Eq.~1, we infer the characteristic radius ($R_{\rm d}$) that would reproduce this precession frequency:

\begin{equation}
  \frac{R_{\rm d}}{a} = \left[ \frac{32}{15} \frac{\sqrt{1+q}}{q} \left( \frac{P_{\rm orb}}{P_{\rm prec}} \right) \right]^{2/3} \approx 0.13.
\end{equation}

\begin{figure*}[h!]
    \centering
    \definecolor{adafL1}{RGB}{255, 245, 190}
    \definecolor{adafL8}{RGB}{130, 10, 5}
    \definecolor{thindisc}{RGB}{85, 105, 130}
    \definecolor{starcore}{RGB}{255, 225, 60}
    \definecolor{staredge}{RGB}{240, 120, 10}
    \definecolor{transred}{RGB}{165, 40, 40}
    \definecolor{jetcol}{RGB}{130, 80, 200}
    \definecolor{jetcore}{RGB}{190, 155, 255}
    \definecolor{jetedge}{RGB}{80, 40, 160}
    \definecolor{adafL5}{RGB}{235, 100, 20}
    \definecolor{adafL7}{RGB}{170, 40, 10}
    \definecolor{earthblue}{RGB}{30, 100, 180}
    \definecolor{earthgreen}{RGB}{50, 140, 60}

    \resizebox{\textwidth}{!}{%
    \begin{tikzpicture}[>=Stealth, font=\small\sffamily]
    
    \draw[black!30, densely dashed, line width=0.6pt] (-7.4,0) -- (8.7,0);
    \node[black!100, font=\tiny\itshape, anchor=north west] at (-7.5, 0.4) {Orbital plane};

    \fill[thindisc!8] (5.5,0) arc (0:180:5.5 and 0.44) -- cycle;
    \fill[white] (2.15,0) arc (0:180:2.15 and 0.15) -- cycle;

    \foreach \rx/\ry/\opf/\opd in {
      5.5/0.44/0.10/55, 5.15/0.41/0.09/45, 4.8/0.38/0.08/38, 4.45/0.35/0.07/32,
      4.1/0.32/0.06/28, 3.75/0.28/0.05/24, 3.4/0.24/0.05/20, 3.05/0.20/0.04/18}{
      \fill[thindisc, opacity=\opf] (\rx,0) arc (0:180:{\rx} and {\ry}) -- cycle;
      \draw[thindisc!\opd, line width=0.3pt] (\rx,0) arc (0:180:{\rx} and {\ry});
    }
    
    \draw[thindisc!70, line width=0.8pt] (5.5,0) arc (0:180:5.5 and 0.44);
    \draw[transred!40, densely dashed, line width=0.8pt] (2.15,0) arc (0:180:2.15 and 0.15);

    \begin{scope}[rotate=14]
        \def\maxR{2.15}
        \def\midH{1.0}
        \def\tipH{0.25}
        \clip (0,0)
            .. controls (0.5*\maxR, \midH) .. (\maxR, \tipH)
            .. controls (\maxR + 0.1, 0) .. (\maxR, -\tipH)
            .. controls (0.5*\maxR, -\midH) .. (0,0)
            .. controls (-0.5*\maxR, \midH) .. (-\maxR, \tipH)
            .. controls (-\maxR - 0.1, 0) .. (-\maxR, -\tipH)
            .. controls (-0.5*\maxR, -\midH) .. (0,0)
            -- cycle;
        \shade[inner color=adafL8, outer color=adafL1!95!white] (0,0) ellipse (2.3 and 1.2);
        \fill[white] (0,0) ellipse (0.34 and 0.11);
        \coordinate (adafTarget) at (-1.0, 0.8);
    \end{scope}

    \begin{scope}[rotate=14]
      \shade[left color=jetcol!20, right color=jetcol!20, middle color=jetcore!40, opacity=0.65] 
        (-0.65,3.1) -- (-0.12,0.10) -- (0.12,0.10) -- (0.65,3.1) arc[start angle=0, end angle=180, x radius=0.65, y radius=0.10] -- cycle;
      \shade[left color=jetcore!10, right color=jetcore!10, middle color=white!70!jetcore, opacity=0.7] 
        (-0.35,2.8) -- (-0.06,0.10) -- (0.06,0.10) -- (0.35,2.8) arc[start angle=0, end angle=180, x radius=0.35, y radius=0.06] -- cycle;
      
      \foreach \y/\rx/\ry/\op in {0.50/0.11/0.035/0.50, 0.95/0.15/0.04/0.42, 1.45/0.19/0.045/0.35, 2.00/0.23/0.05/0.28, 2.60/0.28/0.055/0.20}{
          \shade[inner color=white!60!jetcore, outer color=jetcol!20, opacity=\op] (0,\y) ellipse [x radius=\rx, y radius=\ry];
      }
      
      \draw[jetedge!50, line width=0.4pt, opacity=0.5] (-0.12,0.10) .. controls (-0.2,0.9) and (-0.35,1.8) .. (-0.65,3.1);
      \draw[jetedge!50, line width=0.4pt, opacity=0.5] (0.12,0.10) .. controls (0.2,0.9) and (0.35,1.8) .. (0.65,3.1);
      
      \shade[left color=jetcol!15, right color=jetcol!15, middle color=jetcore!30, opacity=0.55] 
        (-0.50,-2.5) -- (-0.10,-0.10) -- (0.10,-0.10) -- (0.50,-2.5) arc[start angle=0, end angle=-180, x radius=0.50, y radius=0.08] -- cycle;
      \shade[left color=jetcore!5, right color=jetcore!5, middle color=white!50!jetcore, opacity=0.6] 
        (-0.25,-2.2) -- (-0.04,-0.10) -- (0.04,-0.10) -- (0.25,-2.2) arc[start angle=0, end angle=-180, x radius=0.25, y radius=0.05] -- cycle;
      
      \foreach \y/\rx/\ry/\op in {-0.45/0.09/0.025/0.32, -0.90/0.12/0.03/0.25, -1.40/0.15/0.035/0.18, -1.95/0.18/0.04/0.13}{
          \shade[inner color=white!50!jetcore, outer color=jetcol!15, opacity=\op] (0,\y) ellipse [x radius=\rx, y radius=\ry];
      }

      \draw[jetedge!40, line width=0.35pt, opacity=0.4] (-0.10,-0.10) .. controls (-0.15,-0.7) and (-0.28,-1.4) .. (-0.50,-2.5);
      \draw[jetedge!40, line width=0.35pt, opacity=0.4] (0.10,-0.10) .. controls (0.15,-0.7) and (0.28,-1.4) .. (0.50,-2.5);
      
      \node[jetcol!85, font=\tiny\itshape, anchor=south west] at (0.65, 2.4) {Jet};
      \node[jetcol!65, font=\tiny\itshape, anchor=north west] at (0.38, -1.3) {Counter-jet};
    \end{scope}

    \fill[black!10] (0,0) circle (0.38);
    \shade[ball color=black!95] (0,0) circle (0.22);
    
    \node[black, font=\scriptsize, fill=white, inner sep=2pt, anchor=west] at (2.285, 0.65) {$\delta$ (tilt angle)};
    \draw[black, -Stealth, line width=0.8pt] (2.35,0) arc[start angle=0, end angle=14, radius=2.5];

    \fill[thindisc!8] (-5.5,0) arc (180:360:5.5 and 0.44) -- (2.15,0) arc (360:180:2.15 and 0.15) -- cycle;

    \foreach \rx/\ry/\opf/\opd in {
      5.5/0.44/0.10/55, 5.15/0.41/0.09/45, 4.8/0.38/0.08/38, 4.45/0.35/0.07/32,
      4.1/0.32/0.06/28, 3.75/0.28/0.05/24, 3.4/0.24/0.05/20, 3.05/0.20/0.04/18}{
      \fill[thindisc, opacity=\opf] (-\rx,0) arc (180:360:{\rx} and {\ry}) -- (2.15,0) arc (360:180:2.15 and 0.15) -- cycle;
      \draw[thindisc!\opd, line width=0.3pt] (-\rx,0) arc (180:360:{\rx} and {\ry});
    }

    \draw[transred!40, densely dashed, line width=0.8pt] (-2.15,0) arc (180:360:2.15 and 0.15);
    \draw[thindisc!70, line width=0.8pt] (-5.5,0) arc (180:360:5.5 and 0.44);

    \draw[transred!80!black, thin] ( 2.15,-0.15) -- ( 2.15,-2.25);
    \draw[transred!80!black, thin] (-2.15,-0.15) -- (-2.15,-2.25);
    \draw[transred!80!black, <->, line width=0.7pt] (-2.15,-2.15) -- (2.15,-2.15);
    \node[transred!80!black, font=\footnotesize\bfseries, fill=white, inner sep=1.5pt] at (0,-2.15) {$R_d \approx 0.13\,a \approx 10^4 \,R_g$};
    \node[transred!80!black, font=\tiny, anchor=north, align=center, fill=white, inner sep=1.5pt] at (0,-2.425) {Thin disc $\leftrightarrow$ hot flow transition};
    
     \node[adafL7, font=\footnotesize\bfseries, anchor=east] at (-5.2, 4.1) {Hot inner flow};
    \node[adafL5!80!black, font=\scriptsize, anchor=north east, align=right] at (-5.2, 3.95) {Moderately thick\\$H/R \sim 0.1-0.5$\\Precesses as a rigid body};
    \draw[adafL5!80!black, thin, ->, shorten >= 2pt] (-5.5, 3) to[out=-90, in=130] (adafTarget);

    \node[thindisc, font=\footnotesize\bfseries, anchor=west] at (8.2, 3.7) {Outer thin disc};
    \node[thindisc!65, font=\scriptsize, anchor=north west, align=left] at (8.2, 3.55) {$H/R \ll 1$\\Aligned with orbital plane};
    \draw[thindisc!50, thin, ->, shorten >= 2pt] (8.4, 2.85) to[out=-90, in=60] (4.8, 0.45);
    
    \shade[inner color=starcore, outer color=staredge] (5.85,0)
        .. controls (6.1, 0.7) and (7.0, 0.9) .. (7.6, 0.9)
        .. controls (8.4, 0.9) and (8.7, 0.4) .. (8.7, 0)
        .. controls (8.7, -0.4) and (8.4, -0.9) .. (7.6, -0.9)
        .. controls (7.0, -0.9) and (6.1, -0.7) .. (5.85, 0);
    \draw[staredge!55, line width=0.6pt] (5.85,0)
        .. controls (6.1, 0.7) and (7.0, 0.9) .. (7.6, 0.9)
        .. controls (8.4, 0.9) and (8.7, 0.4) .. (8.7, 0)
        .. controls (8.7, -0.4) and (8.4, -0.9) .. (7.6, -0.9)
        .. controls (7.0, -0.9) and (6.1, -0.7) .. (5.85, 0);

    \node[black!70, font=\footnotesize\bfseries] at (7.6,0) {K};
    
    \fill[staredge!80] (5.85,0) circle (0.04);
    \node[staredge!85, font=\tiny, anchor=south east, inner sep=2pt] at (5.85, 0.05) {L$_1$};
    
    \node[staredge!80, font=\scriptsize, anchor=north west] at (8.8,0.60) {Donor star};
    \node[staredge!60, font=\scriptsize, anchor=north west] at (8.8,0.25) {$M_2\!\approx\!0.4\,M_\odot$};
    
    \draw[staredge!80, line width=1.5pt, -{Stealth[length=5pt]}] (5.82,0) .. controls (5.5,-0.5) and (4.7,-0.4) .. (4.1,-0.29);
    \node[staredge!80, font=\tiny\itshape, anchor=north] at (4.8,-0.45) {Mass transfer stream};

    \draw[black!40, densely dashed, line width=0.6pt] (36:2.8) -- (36:6.5);

    \begin{scope}[shift={(36:7.1)}, rotate=36]
        \shade[inner color=earthblue!60!white, outer color=earthblue!80!black] (0,0) circle (0.4);
        
        \begin{scope}
            \clip (0,0) circle (0.4);
            \fill[earthgreen!85!black] (-0.35, 0.15) to[out=-10, in=140] (0.0, -0.05) to[out=-40, in=90] (0.25, -0.25) to[out=180, in=-20] (-0.1, -0.35) to[out=90, in=-110] (-0.35, 0.15) -- cycle;
            \fill[earthgreen!85!black] (0.2, 0.2) circle (0.15);
        \end{scope}
        
        \begin{scope}
             \clip (0,0) circle (0.4);
             \draw[white, opacity=0.5, line width=1.8pt, line cap=round] (-0.3, 0.1) to[out=20, in=160] (0.3, 0.25);
             \draw[white, opacity=0.4, line width=1.2pt, line cap=round] (-0.2, -0.2) to[out=10, in=170] (0.25, -0.1);
        \end{scope}

        \draw[earthblue!30, opacity=0.6, line width=2.5pt] (0,0) circle (0.41);

        \node[black!90, font=\small\sffamily, anchor=south, inner sep=2pt] at (0.35, 0.5) {Earth};
    \end{scope}

    \end{tikzpicture}
    }
    \caption{Schematic view of the geometry considered for A0620--00 in the nodal-precession scenario discussed in this work (not to scale). The outer thin disc (blue) is aligned with the orbital plane. The hot inner flow (orange-red gradient) is moderately thick $(H/R \sim 0.1-0.5)$, misaligned by an angle $\delta$, and precesses as a rigid body. Collimated relativistic jets emerge perpendicular to the hot inner flow midplane. The donor star is shown filling its Roche lobe. The observer is shown at an inclination of $\sim 50^\circ$ relative to the orbital axis \citep{vanGrunsven2017MNRAS.472.1907V}.}
    \label{fig:model_scheme_final}
\end{figure*}
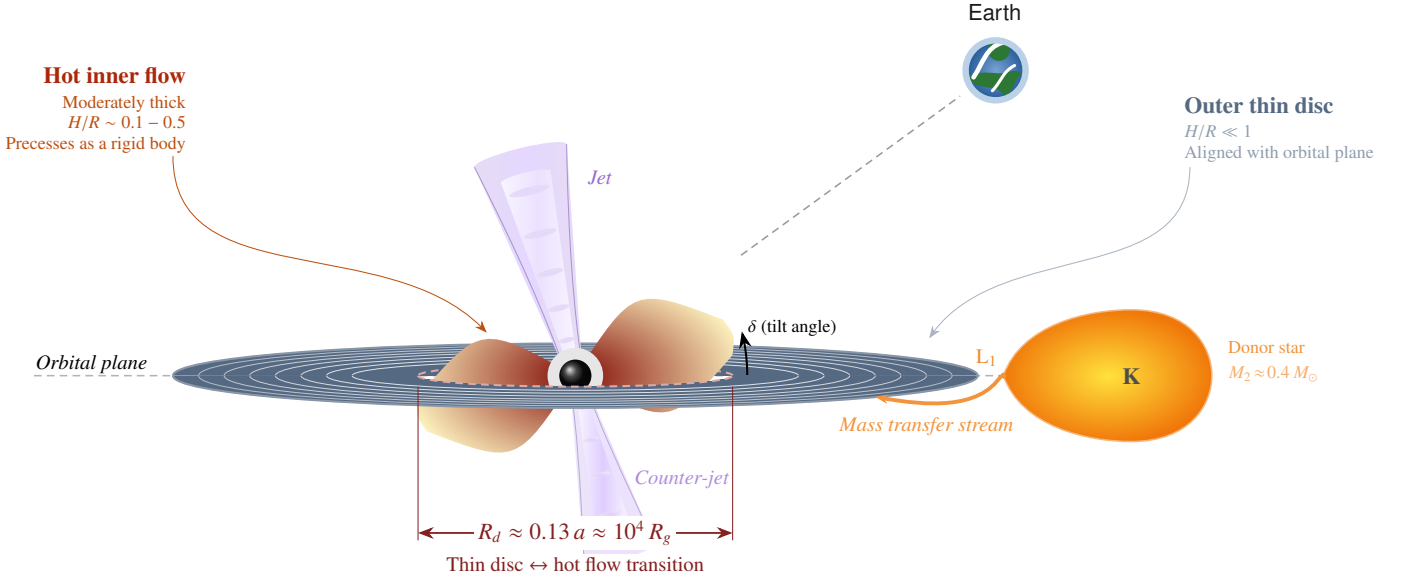

\noindent This derived radius ($R_{\rm d} \approx 0.13\,a \approx 3.6 \times 10^4 R_{\rm g}$) locates the precessing structure deep within the disc, well inside the expected outer disc edge ($R_{\rm out} \sim 0.7\,R_{\rm L1}$, where $R_{\rm L1}$ denotes the effective Roche lobe radius of the compact object; \citealt{Frank2002apa..book.....F}).
A possible solution involves a transition radius between the outer cold, thin disc and the hot inner flow in quiescent BH-LMXBs \citep[$\sim10^4~R_{\rm g}$; see e.g.,][see Fig.~\ref{fig:model_scheme_final} for a schematic representation]{Narayan1996ApJ...457..821N, Esin1997ApJ...489..865E, McClintock2000ApJ...531..956M, Froning2011ApJ...743...26F, Yuan2014ARA&A..52..529Y}.

In quiescence, the accretion flow is likely stratified into two distinct regions: an outer geometrically thin disc and an inner hot, moderately thick flow \citep[e.g.,][]{Narayan1996ApJ...457..821N, Esin1997ApJ...489..865E}. If a misalignment is present, the ability of the flow to precess coherently depends on how efficiently this tilt, or warp, is communicated between neighbouring radii. This can proceed in two limiting regimes: through bending waves when $\alpha \lesssim H/R$, or diffusively when $\alpha \gtrsim H/R$, where $H$ is the local vertical scale height of the accretion flow and $R$ is the cylindrical radius measured in the plane of the disc \citep{Papaloizou1983MNRAS.202.1181P, Papaloizou1995ApJ...438..841P, Lubow2002MNRAS.337..706L}. In the outer thin disc, where $H/R$ is small \citep[$H/R \ll 1$;][]{Shakura1973A&A....24..337S}, the diffusive regime is likely to dominate, at least for commonly adopted quiescent-disc viscosities \citep[e.g.][]{Dubus2001A&A...373..251D, Meyer-Hofmeister2001A&A...372..508M}. This would make it more difficult to sustain a coherent global response to tidal perturbations over large radii. We note that in the aforementioned case of \object{Her~X-1} the situation is significantly different because it is a persistent system characterised by an irradiated, optically thick, geometrically thin disc. In such luminous, irradiated discs, warps can be maintained by external torques, such as radiative and stream-impact torques. The companion tidal torque can then drive coherent precession over larger radii \citep[e.g.,][]{Wijers1999MNRAS.308..207W, Kosec2022ApJ...936..185K}.

As discussed above, for the case of A0620 the precession radius derived for the quiescent superorbital modulation is consistent with the expected transition radius between the outer thin disc and the hot inner flow. For the precession scenario to produce observable photometric modulation, the inner accretion structure should retain a disc-like morphology while remaining thick enough to support coherent bending-wave communication, which requires $\alpha \lesssim H/R$. At the inferred precession radius, the radial sound-crossing time is expected to be much shorter than the precession period, allowing such a compact inner accretion structure to remain radially coupled and precess approximately as a coherent body \citep{Larwood1996MNRAS.282..597L, Larwood1998MNRAS.299L..32L}. We therefore refer to this component as a hot inner flow, a structure with ADAF-like thermodynamic properties but with a moderately thick, disc-like geometry with $H/R$ in the range of $\sim0.1-0.5$ (e.g., \citealt{Ingram2009MNRAS.397L.101I, Das2013MNRAS.435.2431D, Yuan2014ARA&A..52..529Y, Marcel2018A&A...615A..57M}; see Fig.~\ref{fig:model_scheme_final}).  Thus, if the observed precession originates within the accretion flow, the hot inner flow is the more plausible precessing component. 

This interpretation also offers a plausible explanation for the observed phase-dependent occurrence of the two quiescent states, with precession remaining active and cyclically modulating the geometric visibility of the hot inner flow. When the orientation of the hot inner flow reduces its projected contribution, the optical emission would be dominated by the donor star and its ellipsoidal modulation, corresponding to the passive state. Conversely, when the hot flow becomes more directly visible, it would contribute a larger fraction of the total optical flux, producing active-state behaviour. If this geometric picture is correct, it should leave distinctive imprints on the wavelength dependence and viewing-angle sensitivity of the modulation.

The weak chromatic dependence of the modulation offers an additional constraint. The semi-amplitudes in the $g$ and $r$ bands are $0.10 \pm 0.01$ and $0.085 \pm 0.009$~mag, respectively, implying a nearly achromatic signal ($A_{\rm g}/A_{\rm r} \approx 1.18$). Such behaviour is broadly consistent with a geometric origin, in which the variability is driven primarily by changes in viewing angle or path length. By contrast, intrinsic accretion-rate fluctuations would generally be expected to produce stronger variability at bluer wavelengths, while synchrotron emission from a jet might enhance the modulation at longer wavelengths \citep{Russell2006MNRAS.371.1334R, Gallo2006MNRAS.370.1351G}. The present data do not exclude these alternatives completely, however. For example, frequency-dependent optical depths within the flow could still introduce mild chromatic effects \citep{Mahadevan1997ApJ...477..585M, Veledina_2013}. Thus, although the measured amplitudes are compatible with the geometric interpretation, broader wavelength coverage extending into the near-infrared or ultraviolet will be required to distinguish conclusively between these possibilities.

A further geometric diagnostic is provided by the system's viewing angle, which governs the amplitude of any projection-driven modulation. 
The inclination of A0620 ($i \approx 50^{\circ}$; \citealt{Cantrell2010ApJ...710.1127C}; \citealt{vanGrunsven2017MNRAS.472.1907V}) is well suited for detecting geometric modulation from a precessing inner structure. A tilted disc-like surface that contributes a fraction $f$ of the total optical flux produces a peak-to-peak fractional variation that scales as $\Delta F / F_{\rm total} \approx 2\,f\,\delta\,\tan i$\footnote{For a flat emitting surface, the projected area scales as $\cos i$. A tilt $\delta$ causes the effective viewing angle to oscillate between $i \pm \delta$. The peak-to-peak change in projected area is $\cos(i-\delta) - \cos(i+\delta) = 2\sin i\,\sin\delta \approx 2\delta\sin i$ for small $\delta$. Normalising by the unperturbed area ($\cos i$) gives a fractional variation of $2\delta\tan i$.}, where $\delta$ is the tilt angle. At this inclination, $\tan i \approx 1.2$, providing appreciable geometric sensitivity while remaining below the regime where self-shadowing or grazing-angle effects would dominate. Since the donor star dominates the optical flux in the passive state, the hot inner flow likely contributes of order one-third of the total optical flux near active-state maximum. Taking this dilution into account, a tilt angle of $\delta \approx 15^{\circ}$ is sufficient to account for the observed $\sim 0.2$\,mag modulation. The same geometric picture naturally predicts that the modulation amplitude should depend on the system viewing angle: in higher-inclination systems ($i \gtrsim 70^{\circ}$), the steeper projection effects could produce stronger variability or even periodic obscuration episodes, whereas in nearly face-on configurations ($i \lesssim 20^{\circ}$) the signal would be largely washed out. 

The geometric scenario also carries implications beyond the optical band.
Because the X-ray emission in quiescence also originates in the hot inner flow, a corresponding superorbital modulation in X-rays would be expected if the optical signal is indeed geometric in nature. However, the quiescent X-ray luminosity of A0620 is extremely low ($L_{\rm X} \sim 10^{30.8}$\,erg\,s$^{-1}$; \citealt{McClintock1995ApJ...442..358M}), placing such a detection well beyond the reach of current instrumentation. Future X-ray missions with improved soft-band sensitivity could provide an independent test of the precession scenario.

Beyond A0620 itself, the geometric interpretation proposed here is consistent with the growing observational evidence for spin-orbit misalignment in BH-LMXBs. Variations in the jet axis observed during outbursts in \object{V404~Cyg} \citep{MillerJones2019Natur.569..374M} and \object{MAXI~J1820+070} \citep{Poutanen2022Sci...375..874P} suggest that tilted inner structures are present in this population \citep[see][for a review]{Fragile2026arXiv260302993F}. The detection of a superorbital modulation in A0620 during deep quiescence suggests that precessing inner flows may remain dynamically active even at luminosities far below outburst. 
However, this geometric interpretation remains to be tested with dedicated simulations to determine whether a misaligned hot inner flow in A0620 can sustain tidal precession and reproduce the observed superorbital modulation.

\subsection{Alternative scenarios}

Several other mechanisms can also produce long-term optical variability. We examine four possibilities for A0620, including apsidal precession, a hierarchical triple configuration, viscous diffusion, and magnetic activity of the donor star. Although each is physically motivated, none provides a clear explanation for the observed modulation without additional assumptions or independent supporting evidence.

\textit{Apsidal precession} can occur in LMXBs when tidal torques from the donor star excite resonant modes in the accretion disc, particularly in systems with mass ratios $q=M_2/M_1 \lesssim 0.33$ \citep{Whitehurst1991MNRAS.249...25W}. With $q \approx 0.06$ \citep{Neilsen2008MNRAS.384..849N}, A0620 satisfies this low-$q$ condition in principle. Using the formalism of \citet{Patterson2003PASP..115.1308P}, the fractional superhump excess can be expressed as 

\begin{equation}
\Delta P^{-1} = \left[(P_{\rm sh} - P_{\rm orb})/P_{\rm orb}\right]^{-1} = \left[0.37\,q/(1+q)^{1/2}\right]^{-1}\eta^{-2.3} - 1, 
\end{equation}

\noindent where $P_{\rm sh}$ is the superhump period and $\eta \approx 0.8$ parameterises the radial extent of the disc \citep{Mineshige1992PASJ...44L..15M}. The corresponding apsidal precession period is $P_{\rm prec}=\left(1/P_{\rm orb} - 1/P_{\rm sh}\right)^{-1}$. Substituting the parameters of A0620 yields $P_{\rm prec}\approx25~\mathrm{d}$, more than an order of magnitude shorter than the observed 262-day modulation. Therefore, although apsidal precession may operate in low-$q$ systems, the standard resonant-disc framework does not reproduce the timescale observed in A0620.

A \textit{hierarchical triple system} is another possibility, especially given the discovery of a tertiary companion in the BH-LMXB \object{V404~Cyg} \citep{Burdge2024Natur.635..316B}. In this scenario, a third body could perturb the inner binary and drive secular eccentricity oscillations through the Kozai--Lidov mechanism \citep[][]{Kozai1962AJ.....67R.579K,Lidov1962P&SS....9..719L}. This mechanism has also been invoked to explain superorbital modulations in other X-ray binaries such as 4U~1820--303 and 47~Tuc~X--9 \citep{Chou2001ApJ...563..934C,Tudor2018MNRAS.476.1889T}. The characteristic Kozai--Lidov timescale is

\begin{equation}
T_{\rm KL} \simeq \frac{16}{30\pi}
\frac{P_3^2}{P_{\rm orb}}
\frac{M_{\rm tot}}{M_3}
(1-e_3^2)^{3/2},
\end{equation}

\noindent where $M_{\rm tot}=M_1+M_2+M_3$, and $P_3$ and $e_3$ are the orbital period and eccentricity of the tertiary companion \citep{Naoz2016ARA&A..54..441N}. For the quiescent magnitude and distance of A0620 \citep{Cantrell2008ApJ...673L.159C,Cantrell2010ApJ...710.1127C}, a companion contributing only $\sim10\%$ of the optical flux would have $M_V\gtrsim10$, consistent with a late-type main-sequence star such as an M2.5V companion ($M_3\sim0.4~M_\odot$; \citealt{Pecaut2013ApJS..208....9P}).
For this fiducial tertiary mass, reproducing the observed $\sim262$ d modulation requires $P_3\approx5.2-6.4$~d for $0\le e_3\le0.5$; these configurations satisfy the standard hierarchical stability criteria \citep{Mardling2001MNRAS.321..398M}.
Using Kepler’s third law, this corresponds to an outer separation of $a_{\rm out}\approx0.11-0.13~\mathrm{AU}$. Given the inner binary separation $a_{\rm in}=0.0176~\mathrm{AU}$ \citep{Gonzalez2014MNRAS.438L..21G}, this implies a very compact configuration with $a_{\rm out}/a_{\rm in}\approx6-7$.
Thus, although a hierarchical triple configuration cannot be ruled out, there is currently no independent evidence for a tertiary companion in A0620. Moreover, while the Kozai--Lidov mechanism can produce cyclic changes in the inner binary geometry, it would not by itself explain the observed association between the periodic signal, the passive/active states, and changes in the accretion-related variability. These considerations make this scenario less compelling.

\textit{Viscous diffusion} through the outer disc can also produce timescales comparable to the observed superorbital period. 
Using the standard $\alpha$-disc scaling \citep{Shakura1973A&A....24..337S, Frank2002apa..book.....F},
$t_{\rm visc} \approx 3.5\times10^5 \alpha^{-4/5} R_{10}^{5/4} m_1^{1/4}\dot{M}_{16}^{-3/10}\ {\rm s}$,
where $R_{10}=R_{\rm out}/10^{10}$ cm, $\dot{M}_{16}=\dot{M}/10^{16}$ g s$^{-1}$, and $m_1$ is the BH mass in solar units. 
For the parameters of A0620 \citep{Marsh1994MNRAS.266..137M, Cantrell2010ApJ...710.1127C, Froning2011ApJ...743...26F}, a 262 d viscous timescale would imply $\alpha\approx0.02$.
This value is consistent with theoretical expectation for quiescent accretion discs, where low ionization levels reduce turbulent activity \citep[e.g.][]{Menou2000MNRAS.314..498M, Dubus2001A&A...373..251D}.
However, the process is not uniform across the disc due to the radial dependence of the timescale, $t_{\rm visc} \propto R^{5/4}$. The outer regions act as a low-pass filter, damping high-frequency fluctuations from the donor while preserving variations on the order of the viscous timescale. While this mechanism can explain the characteristic duration of the phase-dependent state occurrence, it does not naturally produce a periodic signal like the one observed at $\sim262$ days. Therefore, this hypothesis appears to be unfavourable.  

Finally, a \textit{magnetic activity cycle} of the K-type donor star could also modulate the mass-transfer rate. The K-type donor star in A0620 is expected to be tidally locked to the binary orbit, giving a rotation period $P_{\rm rot} = P_{\rm orb} = 0.323$\,d. In chromospherically active stars, magnetic cycles scale with the stellar rotation period through a ratio that depends on the Rossby number, with values spanning $P_{\rm cyc}/P_{\rm rot} \sim 100$--$1000$ \citep{Brandenburg2017ApJ...845...79B, Strugarek2017Sci...357..185S, Kollath2009A&A...501..695K, Saar1999ApJ...524..295S}. For A0620, the observed ratio $P_{\rm sup}/P_{\rm rot} = 262/0.323 \approx 810$ is consistent within this range. A plausible pathway by which such a cycle could imprint itself on the light curve is through modulation of the mass transfer rate. Cyclic changes in the coverage and distribution of starspots on the donor surface would alter the local effective temperature near the inner Lagrange point, shifting the effective Roche-lobe filling factor and thereby driving quasi-periodic variations in $\dot{M}$ \citep{King2007MNRAS.376.1740K}. In addition, variations in magnetic pressure at the stellar surface could modulate the geometry of the mass transfer stream itself. 

However, the empirical activity--rotation relations have been calibrated primarily on slowly rotating, partly convective solar-type stars, and their extrapolation to the rapidly rotating, fully or nearly convective donor in A0620 is uncertain. It is also unclear whether a magnetic cycle in such a star would produce a sufficiently coherent and large-amplitude signal to account for the $\sim$0.2\,mag modulation observed here, given that the resulting $\dot{M}$ variations would still need to propagate through the disc before affecting the optical flux. This mechanism cannot be excluded with the present data, and could be tested through long-term spectroscopic monitoring of chromospheric activity indicators such as H$\alpha$ \citep{Gonzalez2010A&A...516A..58G}, which would track the donor's magnetic state independently of the accretion flow.

\section{Conclusion}

Using optical photometry spanning nearly two decades, we analysed the quiescent black-hole X-ray binary A0620--00. The data reveal a statistically robust superorbital modulation at $P=261.9\pm9.4$~d that is unlikely to arise solely from instrumental effects or sampling artefacts. The observed periodicity is plausibly interpreted as the signature of retrograde nodal precession of the inner accretion flow. The corresponding dynamical radius of $\sim0.13a$ places the precessing structure well within the outer disc edge, consistent with the expected transition radius between the outer thin disc and the hot inner accretion flow. 

These results have direct implications for how accretion dynamics are interpreted in quiescent BH-LMXBs. If the observed modulation is indeed driven by nodal precession, then the inner accretion flow is not a static structure, but one with persistent behaviour whose cyclic geometric reorientation regulates the visibility of the inner region in the optical flux. This offers a plausible explanation for the phase-dependent occurrence of the active and passive states without necessarily invoking large changes in the mass transfer rate. The broader implications are also significant. If similar superorbital signals are present but remain undetected in other systems, then precessing inner flows may represent a fundamental, yet so far overlooked, feature of low-luminosity accretion.

High-cadence optical monitoring with next-generation facilities such as the Vera~C.\ Rubin Observatory (and its LSST) will provide an important opportunity to determine whether similar superorbital signals are common among quiescent BH-LMXBs. 
At the same time, dedicated numerical simulations will be needed to test whether a misaligned hot inner flow can sustain coherent tidal precession and generate a long-period modulation comparable to that observed in A0620.

\begin{acknowledgements}
We thank the anonymous referee for their comments, which have improved the manuscript. 
We acknowledge support by Spanish Agencia Estatal de Investigación via PID2021-124879NB-I00, PID2022-143331NB-100 and PID2024-161863NB-I00.
D.M.S. and M.A.P. acknowledge support through the Ramón y Cajal grants RYC2023-044941 and RYC2022-035388-I, funded by MCIU/AEI/10.13039/501100011033 and FSE+.
DMR is supported by Tamkeen under the NYU Abu Dhabi Research Institute grant CASS.
Funded by the European Union (Project 101183150 -- OCEANS). Views and opinions expressed are, however, those of the author(s) only and do not necessarily reflect those of the European Union or the European Research Executive Agency (REA). Neither the European Union nor REA can be held responsible for them. This work makes use of observations from the Las Cumbres Observatory global telescope network.
Based on observations obtained with the Samuel Oschin Telescope 48-inch and the 60-inch Telescope at the Palomar
Observatory as part of the Zwicky Transient Facility project. ZTF is supported by the National Science Foundation under Grants
No. AST-1440341 and AST-2034437 and a collaboration including current partners Caltech, IPAC, the Oskar Klein Center at
Stockholm University, the University of Maryland, University of California, Berkeley, the University of Wisconsin at Milwaukee,
University of Warwick, Ruhr University, Cornell University, Northwestern University and Drexel University. Operations are
conducted by COO, IPAC, and UW.
\end{acknowledgements}

\bibliographystyle{aa}
\bibliography{biblio}

\end{document}